\documentclass{emulateapj} 

\usepackage{graphicx,amsmath}

\slugcomment{ApJ, accepted}

\shorttitle{Aberrations in Lyot Coronagraphy}
\shortauthors{Sivaramakrishnan et al.}

\newcommand \eg     {{\it e.g., }}

\newcommand \ie     {{\it i.e.,}}
\newcommand \viz    {{\it viz.,}}

\newcommand \eq     {\,=\,}                 
\newcommand \sinc    {{\rm sinc}}
\newcommand \saw    {{\rm \Lambda}}
\newcommand \conv    {{\rm *}}
\newcommand \bx    {{\bf x}}
\newcommand \bk    {{\bf k}}

\begin{document} 

\title{Low Order Aberrations in\\
       Band-Limited Lyot Coronagraphs }

\author{Anand Sivaramakrishnan\altaffilmark{1}, R\'emi Soummer\altaffilmark{1,2}}
	\affil{Space Telescope Science Institute\\
		3700 San Martin Drive, Baltimore, MD 21218}

\author{Allic V.\ Sivaramakrishnan}
	\affil{Choate Rosemary Hall School\\
		333 Christian Street, Wallingford, CT 06492}
\author{James P. Lloyd\altaffilmark{1}}
	\affil{Astronomy Department\\
		 230 Space Sciences Building \\
		Cornell University, Ithica, NY 14853}

\author{Ben R.\ Oppenheimer}
        \affil{Astrophysics Department, American Museum of Natural History\\
                Central Park West at 79th Street, New York, NY 10024}
   
			\and
\author{Russell B.\ Makidon\altaffilmark{1}}
	\affil{Space Telescope Science Institute\\
		3700 San Martin Drive, Baltimore, MD 21218}

\altaffiltext{1}{NSF Center for Adaptive Optics.}
\altaffiltext{2}{Michelson Post-doctoral Fellow.}

	\newpage
\begin{abstract} 
We study the way Lyot coronagraphs with unapodized entrance pupils
 respond to small, low order phase aberrations.  This study is applicable to
 ground-based adaptive optics coronagraphs operating at 90\% and higher
 Strehl ratios, as well as to some space-based coronagraphs with intrinsically
 higher Strehl ratio imaging.  
We utilize a second order expansion of the monochromatic point-spread function
 (written as a power spectrum of a power series in the phase aberration
 over clear aperture) to derive analytical expressions for the response of a 
 `band-limited'  Lyot coronagraph (BLC) to small, low order,  phase aberrations.
The BLC possesses a focal plane mask with an occulting 
spot whose opacity profile
is a spatially band-limited function rather than a hard-edged, opaque disk.
The BLC is, to first order, insensitive to tilt and astigmatism.
Undersizing the stop in the re-imaged pupil plane (the Lyot plane) 
following the focal plane mask can alleviate second order effects of astigmatism,
at the expense of system throughput and angular resolution.
The optimal degree of such undersizing depends on individual
instrument designs and goals. Our analytical work
engenders physical insight, and complements existing
numerical work on this subject.
Our methods can be extended to treat the passage of higher order
aberrations through band-limited Lyot coronagraphs, by using our polynomial
decomposition or an analogous Fourier approach.
\end{abstract}


\section{Introduction}
Lyot coronagraphy \citep{Lyo30, Lyo39} has enjoyed a resurgence because interest in
discovering and characterizing
extrasolar planets has been stimulated by advances
in ground-based adaptive optics (AO), as well as support from space agencies
for extrasolar planetary science missions.
There is currently one extreme adaptive optics (ExAO) coronagraph being used to
conduct a complete complete survey for faint companions and disks around nearby stars 
\citep{OSM03, OppenheimerSPIE04, DigbySPIE04, MSP05}.
New coronagraph designs to enable the detection and characterization
of extrasolar planets abound
\citep{NP01, ASF02, Kuchner02, SAF03, SDA03, Aime03bb, KVS03, Soummer05}.
Many of these designs would suppress light from the central star around
which a planet orbits, as long as the telescope and instrument optics
and stops were perfect, and simple Fourier optics theory was accurate enough
to predict instrument behavior in these high contrast regimes.
Less-than-ideal image quality is a stumbling
block which impedes attainment of this science goal.
Kasdin (private communication) has demonstrated in the laboratory that Fourier optics 
appears to be valid down to contrast ratios of $10^{-7}$.

We study the way Lyot coronagraphs dedicated to imaging extrasolar 
Jovian and terrestrial planets respond to small, low order phase aberrations.
\citet{Perrin03} demonstrated that for ground-based adaptive optics systems
delivering Strehl ratios of at least 90\%, a second-order expansion of the 
imaging system's point-spread function (PSF) response to
monochromatic light from a point source \citep{Sivaramakrishnan02}
was sufficient to model the effects of phase aberrations on the PSF.
We analyze the case of unapodized telescopes in high dynamic range regimes, as is relevant
to the problem of tolerancing optics on space-based coronagraphic telescopes or 
calibrating wavefront sensing non-common path errors on ground-based ExAO
instruments dedicated to finding and characterizing extrasolar planets.

\citet{Lloyd05} derived the response of a simple band-limited
coronagraph to small tilt errors, using an analytical method combined with
an expansion of the pupil plane field strength (defined later) in terms of the 
phase aberration over the pupil.  
That work produced insight into the way tilt error causes light to leak
through a coronagraph that is designed to suppress all light
from a perfectly flat, on-axis wavefront. 
Such an analytical treatment develops a qualitative understanding
of coronagraphic response to phase errors.  Insights developed there
apply to classical Lyot coronagraphy.
We extend the methods of \citet{Lloyd05} to understand how
low order aberrations propagate through a band-limited Lyot coronagraph.
While earlier studies \citep[\eg][]{Malbet96,Sivaramakrishnan01,Green03}
explored some aspects of coronagraphy on imperfect wavefronts, they did not 
provide analytical tools required to understand some
of the optical problems of designing a coronagraph to achieve
the $10^7$ photometric dynamic range needed by a space mission dedicated
to discovering and characterizing extrasolar Jovian planets,
or the $10^{10}$ contrast ratio that extrasolar terrestrial
planet discovery may well demand \citep{Breckinridge04}.
Our analytical approach is a start along the route to understanding
coronagraphic data described in \eg\ \citet{OppenheimerSPIE04, DigbySPIE04}.

A band-limited coronagraph is a perfect coronagraph design in that
simple Fourier optics modelling suggests that it will prevent all incoming,
on-axis light from reaching the final coronagraphic focal plane.
We define light leak as the fraction of incident on-axis energy that
reaches the final coronagraphic focal plane. 
Our concern is restricted to diffractive light leak rather than light
redistributed within the instrument by scattering off imperfect optics.
We also concentrate on small aberrations of the wavefront that forms
the first image. For instance, we treat tilts that are small compared to the angular
resolution of the system.  Such tilts
decenter the star behind the focal plane mask's occulting spot without moving it out from
behind the spot ---  which is itself at least a few resolution elements in diameter.
Such aberrations in the wavefront prior to the focal plane mask result in light
leaking through a supposedly perfect coronagraph, reaching the final
coronagraphic image plane, thereby reducing the dynamic range of the instrument.
We find that small amounts of tilt and astigmatism do not
increase the light leak through the coronagraph, whereas the same
amounts of coma and spherical aberration do.  However, detrimental
second order effects of astigmatism can be reduced by undersizing
the Lyot stop.
Our analysis shows that defocus and tilt produce characteristic intensity
distributions in the Lyot plane, which suggests ways this signal can be used to 
align a high dynamic range coronagraph, and calibrate the 
offset between the focal plane mask location in the final coronagraphic
image plane, thereby improving the coronagraph's astrometric accuracy.
Similar analyses on the various coronagraph designs being
developed for the purpose of detecting Jovian and Earth-like planets around
nearby stars will contribute to an informed selection of a design, 
in addition to providing insight into how to improve these designs.

\section{Basic Coronagraphic Theory}

A simple mathematical description of the fundamentals of Lyot
coronagraphy \citep{Lyo39} can be found in \citet{Sivaramakrishnan01},
\citet{AS02b},  \citet{Lloyd05}, and references therein. 

\subsection{Monochromatic coronagraphic theory}

Here we briefly recapitulate our basic monochromatic Fourier optics formalism,
which follows that of \citet{Lloyd05, SL05}.
A more detailed treatment can be found in \citet{Born93}.
We recollect that a plane monochromatic wave
travelling in the $z$ direction in a homogenous medium without
loss of energy can be characterized by a complex
amplitude $E$ representing the transverse (\eg\ electric) field strength of the wave.
The full spatio-temporal expression for the field strength is
$E e^{(i\kappa z - \omega t)}$,
where $\omega/\kappa  = c$, the speed of the wave.
We do not use the term {\it field} to denote image planes --- 
the traditional optics usage --- we always use the term to denote
electromagnetic fields or scalar simplifications of them.
The wavelength of the wave is $\lambda = 2\pi/\kappa $.
The time-averaged intensity of a wave at a point is proportional to $EE^*$,
where the average is taken over one period,  $T = 2\pi/\omega$,
of the harmonic wave.
The phase of the complex number $E$ represents a phase difference from the
reference phase associated with the wave.   A real, positive $E$ corresponds to 
an electric field oscillating in phase with our reference wave.  A purely imaginary
positive value of $E$ indicates that the electric field lags by a quarter cycle from the 
reference travelling wave.
Transmission through passive, linear filters such as apertures, 
apodizers, and so forth, is represented by multiplying the field strength by the
transmission of these objects which modify the wave.  Again, such multiplicative
modification is accomplished using complex numbers to represent 
phase changes forced on the wave incident on such objects.

We assume that scalar Fourier optics describes our imaging system
\citep[cf.][]{Breckinridge04}: image field strengths are given by the Fourier transform of 
aperture (or pupil --- we use the two terms interchangeably) illumination functions, and vice versa.

\begin{figure}
	\plotone{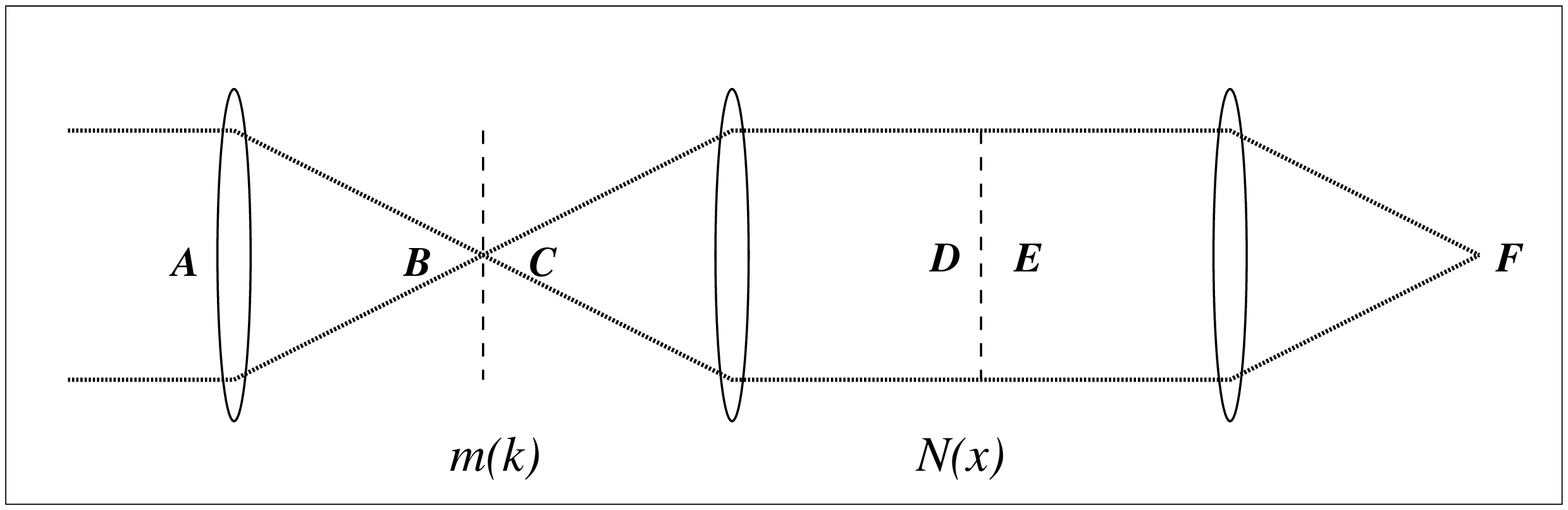}
	\caption{
		The essential planes and stops in a coronagraph.  The entrance aperture
		is A, the direct image at B falls on a focal plane mask whose transmission function
		is $m(k)$. The re-imaged pupil plane D, after being modified by passage through
		a Lyot stop with a transmission function $N(x)$, is sent to the coronagraphic
		image at F.  A, D, and E are pupil planes, and B, C, and F are image planes.
	} 
\label{fig:corolayout}
\end{figure}

\begin{figure}
	 \plotone{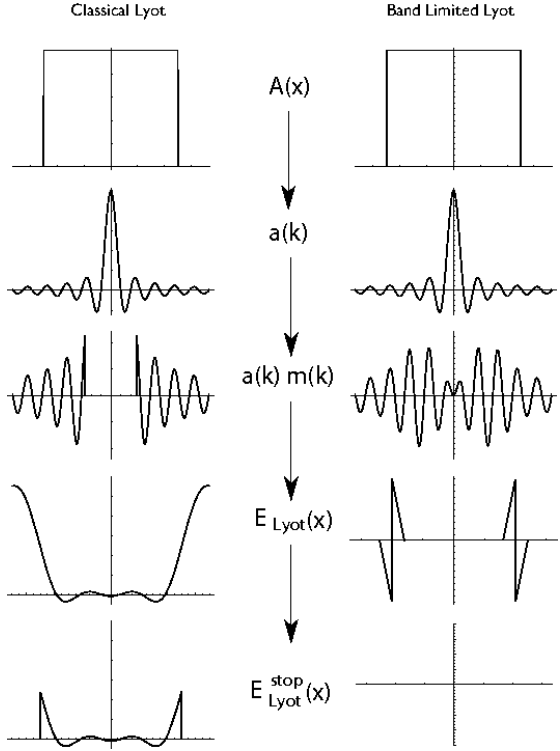}
	 \caption{
		A sketch of the classical Lyot coronagraph (left), and the 
		$1 - {\rm sinc}$ ``top-hat'' band-limited coronagraph operating
		on a perfectly flat incoming wavefront.
		The top row shows the same aperture function used in both 
		examples, \viz\  a clear unobscured entrance aperture.
		The second row shows the field strength at the image plane.
		The third row shows the field strength immediately
		after passage through the focal plane mask.  The classical
		Lyot, on the left, has a hard-edged, opaque mask.  
		This band-limited
		coronagraph has a mask that is opaque only at its center.
		The fourth row shows the field strength in the Lyot pupil 
		plane before a Lyot stop is applied, and the final row
		shows the field after the Lyot stop in the Lyot pupil plane.
		The band-limited coronagraph blocks all incoming 
		on-axis light if the wavefront is unaberrated.
	}
	\label{fig1}
 \end{figure}

A telescope aperture is described by a transmission function
pattern $A(\bx)$, where $\bx=(x,y)$ is the location in
the aperture, in units of the wavelength of the light (see Fig.~\ref{fig:corolayout}).
The corresponding aperture illumination describing the electric
field strength in the pupil (in response to an unaberrated,
unit field strength, monochromatic incident wave)  is $ E_A =  A(\bx) $.
From this point onwards we drop the common factor $E e^{(i\kappa z - \omega t)}$
when describing fields.  The aperture intensities ($E_A E_A^*$) for two
coronagraphic designs are shown in Fig.~2 (top row).
The field strength in the image plane, $E_B = a(\bk)$,
is the Fourier transform of $E_A$, where
$\bk=(k_x,k_y)$ is the image plane coordinate in radians.
Because of the Fourier relationship between pupil and image fields,
$\bk$ is also a spatial frequency vector for a given wavelength of light.
We refer to this complex-valued field $a$ as the `amplitude-spread function' (ASF),
by analogy with the PSF of an optical system. The PSF is $aa^*$.
Our convention is to change the case of a function to indicate its Fourier
transform. We multiply  the image field $E_B$ by a mask function $m(\bk)$ to
model the focal plane mask of a coronagraph.
The image field immediately after this mask is $ E_C = m(\bk) \,  E_B $.
The electric field in the re-imaged pupil after the focal plane mask 
is $E_D$, which is the Fourier transform of $E_C$. 
We use the fact that the transform of the image plane field $E_B$
is just the aperture illumination function $E_A$ itself, so the Lyot pupil field is
$ E_D =  M(\bx) \conv  E_A $, where  $\conv$ is the convolution operator.

If the Lyot pupil stop transmission is $N(\bx)$, the electric
field after the Lyot stop is $E_E = N(\bx) E_D$.  The transform of this
last expression is the final coronagraphic image field strength:
$ E_F  = n(\bk) \conv [m(\bk) \, E_B]$.

The aperture illumination function with phase aberrations is 
\begin{equation} \label{Aaber1}
A_{aber} \eq  A(\bx) e^{i\phi(\bx)}.
\end{equation}
In this paper we look at the way small phase aberrations $\phi(\bx)$
pass through a band-limited coronagraph.  Our approach is to 
expand the exponential in equation (\ref{Aaber1}) in powers of $\phi(\bx)$,
so the quantity that determines the rate of convergence of the expansion 
is the largest excursion of $\phi$ from its mean value over the clear aperture.

\subsection{The Lyot plane field}

A coronagraphic stop is modelled by multiplying the image field by
a focal plane mask transmission function $m(\bk)$,
so the electric field strength just after the focal plane mask is $a_{aber} m$.
When describing a Lyot coronagraph it is sometimes helpful to 
introduce a `mask shape function' $w$ by the definition
\begin{equation} \label{maskfunction}
m(\bk) \equiv  1 - w(\bk).
\end{equation}
When $w$ is unity at the origin  the mask is opaque at its center.
We can  write the electric field at the Lyot plane as
\begin{equation} \label{lyotstopfield}
\begin{array}{ll}
 E_{Lyot}(\bx) & \eq  A_{aber} \conv (\delta(\bx) - W(\bx)) \\
    	              & \eq  A_{aber} - A_{aber} \conv W(\bx).
 \end{array}
\end{equation}
Understanding the subtleties of this equation in the case
of high Strehl ratio imaging is essential to understanding
how small phase aberrations cause light to leak through a
coronagraph designed to produce perfect on-axis image
suppression with completely unaberrated on-axis light.

\subsection{Band-limited coronagraphs} 

\citet{Kuchner02} use a focal plane mask shape function $w$ which is band-limited.
This means that there is a minimum positive value of $b$ such that
the mask function's FT, $W$, satisfies the property 
\begin{equation} \label{bandlimit}
W(\bx) \eq 0  {\rm \ if\ } |\bx| > b.
\end{equation}
The bandwidth or bandpass of $w$ is $b$ (note that $b$ is actually a physical
distance in pupil space). We select the mask function in its transform (pupil) space, $(x,y)$,
rather than in physical (image) space, $(k_x, k_y)$, even
though it is applied in the image plane in any real coronagraph.
If the telescope diameter is $D$, then the characteristic
scale of the mask function is $D/b$ resolution elements
(a resolution element is $\lambda/D$ radians, $\lambda$ 
being the wavelength of the monochromatic light).
This results in a focal plane mask about $D/b$ Airy rings in size.

\subsection{The top-hat band-limited coronagraph} 

	The simplest choice for $W(\bx)$ is the `top-hat' function
	$\Pi(x/d, y/d) / d^2 $, where 
	\begin{equation} 
    	\begin{array}{ll} \Pi(x,y) \eq 1 & {\rm for\ } |x| < 1/2, |y| < 1/2, \\
                      \Pi(x,y) \eq 0   & {\rm elsewhere}
    	\end{array}
    	\end{equation}
	(see Fig.~\ref{fig1}).
	The bandwidth of the corresponding mask function is $d/2$.
	A normalizing factor of $1/d^2$ is applied to ensure that the area
	under $W(\bx)$ is unity, thus ensuring that $w(0,0) = 1$; 
	the focal plane mask is opaque at its center.
	A top-hat $W$ produces a multiplicative image stop function
	\begin{equation} 
	m(\bk) = 1 - \sinc(d k_x)\, \sinc(d k_y).
    	\end{equation}
	Our interest such a choice is primarily didactic:
	this coronagraph leaks aberrated light in a manner
	similar to that of the more popular
	`sawtooth' band-limited coronagraph, \viz\ one with a 
	mask function FT of
	\begin{equation} 
	W(x,y) = \saw(x/d, y/d) / d^2 \eq \saw(x/d) \saw(y/d) / d^2,
    	\end{equation}
	where
	\begin{equation} 
    	\begin{array}{ll} \saw(x) \eq |1 - x| & {\rm for\ } |x| < 1, \\
                      \saw(x) \eq 0   & {\rm elsewhere}.
    	\end{array}
    	\end{equation}
	The sawtooth coronagraph has a focal plane mask described by
	$m(\bk) = 1 - \sinc^2(d k_x)\, \sinc^2(d k_y)$.

\section{Expansion using the Phase Function}

	The phasor $e^{i\phi}$ can be expanded as $1 + i\phi - \phi^2/2! + ...$ for any finite value of 
	the aberration $\phi$. Depending on the size of the aberration it is useful to truncate
	this expansion at various orders \citep[\eg][]{Bloemhof01,Sivaramakrishnan02, Perrin03}.
	The first order perturbation of the field strength in the aperture is imaginary: \ie\ the field due to
	phase aberrations is in phase-quadrature to the perfect aperture field.  The second order aberration of the
	aperture field strength is in (anti-)phase with the perfect wave's field strength.

\subsection{Zernike polynomials and Cartesian expansions}

	Zernike polynomials, which are a set of orthogonal polynomials
	on the unit radius 2-dimensional disk, are frequently used to describe
	phase aberrations in optical systems \citep[\eg][]{Born93, Noll76}.
	These polynomials are expressible in either circular coordinates
	$(r,\theta)$ or Cartesian coordinates $(x,y)$. Expressing Zernike polynomials in 
	Cartesian coordinates leads to a mathematical simplicity
	that promotes a better understanding of the physics
	of an imperfect band-limited Lyot coronagraph.  
	Because of this, we investigate coronagraphy on a
	square aperture in order to make integrals separable in $x$ and $y$.

	We have the choice of of
	writing the phase as either a simple polynomial or an expansion in
	terms of Zernikes.  In the latter case, we must choose the Zernike
	functions' normalized coordinates: we choose the disk on which we define our
	Zernike functions to possess an area equal to that of our square
	aperture.  This choice results in extrapolating the Zernike polynomials
	to a radial coordinate of $2/\sqrt \pi \simeq 1.128$, which is beyond their
	usual range of validity.   Doing so	is a fruitful exercise
	if we restrict ourselves to low order aberrations.
	We emphasize that we do not invoke any of the Zernike polynomials' orthonormal
	properties to prove any result.
	Thus our apparently cavalier extrapolation of the polynomials beyond the
	domain over which they are usually defined does not lead to grief. 
	We only use them here because they are widely used in the optical literature.
	A more natural Fourier decomposition of the phase aberration over the
	aperture is of course possible, and may even preferrable to a Zernike
	decomposition for higher order aberrations.

	Our choice of the equal-area circle for our Zernike polynomials
	is motivated by the following considerations:  if we use the circumscribing
	circle around our square aperture, we miss the steep increases of 
	Zernike functions describing \eg\ spherical aberration at the
	edge of the disk, which for the most part will lie outside our
	square aperture.  If we use an inscribed circle, we include large areas in the
	corners of the apertures where the magnitude of the third, fourth, and
	fifth order polynomials increase sharply, contributing
	features that we do not commonly associate with particular
	Zernike aberrations.  Instead of either extreme, we stick to the middle ground,
	at the expense of mathematical purity: we miss some of the higher values
	of these Zernike polynomials where our circle extends outside the square aperture,
	but collect larger phase errors at the corners of our aperture, where
	the normalized Zernike arguments exceed unity. The net effect of this is that
	we can still talk about Zernike functions on our square aperture,
	in much the same way Zernike function aberrations on obscured
	apertures such as HST or Palomar are discussed, even though
	the annular Zernike functions of \citet{Mahajan81} should be used on annular apertures.

  \section{Propagation of Simple Polynomial Phase Aberrations}

	At any location in the pupil plane, $A_{aber}$  can be expanded
	in an absolutely convergent series in $\phi$ for any finite value of the phase function:
\begin{equation} \label{lyot_expansion}
     A_{aber} \eq  A A_{\phi} \eq  A ( 1 + i\phi  -  \phi^2/2 + ...).
    \end{equation}
	If the phase over the aperture is expressed by
\begin{equation}  \label{zpoly}
	\begin{array}{ll} 
		\phi (x,y) & \eq \displaystyle \sum_{i=1}^{\infty} a_{i} Z_i(x,y)  \\
		           & \eq  \displaystyle \sum_{n=0, m=0}^{\infty} \alpha_{mn} x^n y^m,
\end{array}
\end{equation}
where the $Z_i$'s are Zernike polynomials, and $a_{i}, \alpha_{mn} $
	are constant coefficients, then equation (\ref{lyot_expansion})
	produces first, second, and higher order terms in $x$ and $y$. A measure of the
	magnitude of the phase aberration ---  $\epsilon$, the largest absolute value of the deviation
	of the phase $\phi$ from its aperture-weighted mean --- can be used 
	to estimate the size of a particular order term in equation~(\ref{lyot_expansion}).
	Given any particular value of $\epsilon$ one can estimate where the terms in the above expansion
	become negligible \citep{Perrin03}. For example, in the Advanced Camera for Surveys (ACS) coronagraph
	on the Hubble Space Telescope (HST), which sits in the aberrated beam,
	with about half a wave of spherical aberration at a wavelength of 800nm,
	one must use about seven terms of the above expansion to model the
	gross features of the aberrated field strength over the aperture.

\bigskip

  \subsection{Power leak through a top-hat band-limited coronagraph} \label{top}

	We can calculate how an arbitrary polynomial phase aberration of the form 
	$\alpha_{nm} x^n y^m$ 
	propagates through the coronagraph to the Lyot plane, using this expansion.
	The calculation is straightforward for a square aperture because
	the limits of integration and the basis functions are separable.
	From equation (\ref{lyotstopfield}) we see that the Lyot field for such an
	aberration is described by
\begin{equation}  \label{lyotexp}
\begin{array}{ll} 
E_{Lyot}(x,y) & \eq   A (x,y) \alpha_{nm} x^n y^m \\
              & - \Pi(x/d,y/d) / d^2 \conv [A(x,y)  \alpha_{nm} x^n y^m] 
\end{array}
\end{equation}
	for the `top-hat' band-limited coronagraph design. This calculation becomes difficult to perform analytically
	when the mask transform function has a more extended support, as in  a classical Lyot design.

	We define the function
\begin{equation}  \label{Lpi}
	L_{\Pi,n}(x,d)  \eq  \Pi(x/d) \conv x^n / d,
\end{equation}
which enables us to write equation (\ref{lyotexp}) as
\begin{equation}  \label{lyotexp1}
\begin{array}{ll} 
E_{Lyot}(x,y) & \eq   \alpha_{nm} x^n y^m \\
              & - \alpha_{nm}  L_{\Pi,n}(x,d) L_{\Pi,m}(y,d)
\end{array}
\end{equation}
	in the interior of the Lyot plane in the case of unapodized apertures.
	We define the interior of the aperture as that part of the aperture
	(or image of the aperture, such as the Lyot plane) which is
	further than $b$ from any point on the aperture boundary (where $b$ is the
	bandwidth of the mask shape function).
	This is the area of the Lyot pupil which would not be obstructed by a Lyot stop in
	a perfect band-limited coronagraph \citep[see][for further detail]{Kuchner02, Lloyd05, SL05}.
	The aperture function $A(\bx)$ is unity everywhere in the interior
	of the aperture for an unapodized entrance pupil, and so does not appear
	explicitly in equation (\ref{lyotexp1}).

	The 2.4~m HST ACS coronagraph has an image plane stop diameter of $1\farcs8$,
	or $\sim 30$ resolution elements at 800~nm. Therefore the results we derive here for
	band-limited coronagraphs apply even though the ACS coronagraph is not band-limited,  because
	its Lyot plane can be said to possess an `interior' area which is further than a few $D/30$ scale lengths
	away from the annular aperture boundaries \citep[see][for a discussion of the natural scale length in the 
	Lyot plane]{Sivaramakrishnan01}.
	
	Explicit evaluation $L_{\Pi,n}(x,d)$ is a matter of straightforward integration: the first few values of $n$ produce
\begin{equation}  \label{Lpi_n}
\begin{array}{ll} 
	1   - L_{\Pi,0}(x,d) & \eq  0                                            \\
	x   - L_{\Pi,1}(x,d) & \eq  0                                            \\
	x^2 - L_{\Pi,2}(x,d) & \eq     -d^2     / 12                              \\
	x^3 - L_{\Pi,3}(x,d) & \eq     -d^2 x   / 4                               \\
	x^4 - L_{\Pi,4}(x,d) & \eq     -d^2 x^2 / 2 -   d^4       / 80            \\
	x^5 - L_{\Pi,5}(x,d) & \eq   -5 d^2 x^3 / 6 -   d^4 x     / 16            \\
	x^6 - L_{\Pi,6}(x,d) & \eq   -5 d^2 x^4 / 4 - 3 d^4 x^2 / 16 - d^6 / 448. \\
\end{array}
\end{equation}
	The right hand sides of these equations are the first order 
	residual Lyot field strength given a small phase aberration of $\alpha_n x^n$
	over the entrance pupil of a one-dimensional band-limited coronagraph.

 \begin{figure}
	 \plotone{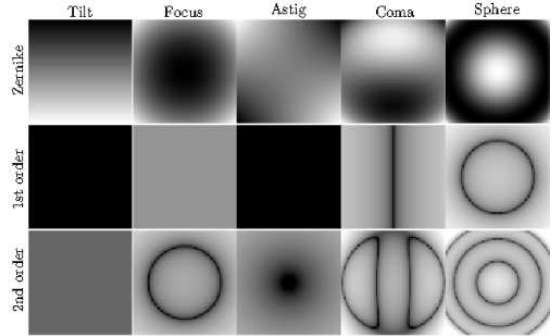}
	 \caption{
		Response of the $1 - {\rm sinc}$ `top-hat' band-limited coronagraph
		to tilt, astigmatism, focus, coma and spherical aberrations. 
		Each aberration shown produces a 95\% Strehl ratio direct image.
		Phase aberrations (top row)  result in non-zero intensity in 
		the Lyot pupil interior.  Analytical first order (middle row)
		and second order (bottom row) approximations to these intensities
		are shown on a logarithmic grey scale, between $0.1$ (white) and 
		$10^{-7}$ (black). 	First order power leak through the BLC due to 
		tilt and astigmatism are identically zero.
		Defocus, and second order tilt cause uniform illumination in the 
		Lyot pupil interior, which produces a fainter version of the
		direct PSF (although with a larger resolution element due to Lyot
		stop undersizing) in the final coronagraphic image.
	}
	\label{killerfig}
 \end{figure}

	Low order phase aberrations, along with
	the resultant first and second order Lyot plane intensities are shown
	in each column of Fig.~\ref{killerfig}.  The band limit of this
	coronagraph is $b = D/5$.  The fraction of incident
 	power leaking through into the Lyot plane interior
	is presented in Table~\ref{table1}.  Each power leak fraction is 
	renormalized by dividing by the fractional
	coronagraphic throughput, which is the ratio of the  areas of 
	the optimal Lyot stop and the entrance aperture.  For a band
	limit of $D/5$ this ratio is $9/25$ \citep[\eg][]{Sivaramakrishnan01}.

\section{Extension to the Sawtooth Band-Limited Coronagraph} \label{saw}

	Designs similar to the sawtooth  (sinc-squared focal plane mask function)
	coronagraph are being considered for the James Webb Space Telescope's
	Near Infrared Camera, as well as for a dedicated coronagraphic space telescope.
 	In order to calculate how the sawtooth band-limited coronagraph leaks low order aberrations, 
	we can define a function $L_{\Lambda,0}(x,d)$ using the analogy of equation (\ref{Lpi}):
\begin{equation}  \label{Llam}
L_{\Lambda,n}(x,d)  \eq  \Lambda(x/d) \conv x^n / d,
\end{equation}
and carry out the same analysis as in section \ref{top}. We present the top-hat band-limited coronagraph's 
	response to low order aberrations (Fig.~\ref{killerfig}); the sawtooth coronagraph's response is qualitatively similar.
	The latter coronagraph's mask function's Fourier transform is a sawtooth function, which also yields
	tractable algebraic expressions for the Lyot pupil field strength for any order.
	For the sawtooth design the residual Lyot field in the interior of the Lyot plane has similar algebraic properties to that of
	the top-hat design. The powers of the relevant physical quantities for the $1 - {\rm sinc}^2$ mask function of the
	sawtooth coronagraph are the same as those in equation (\ref{Lpi_n}), although the numerical coefficients differ slightly.

\section{ Secondary Mirror Alignment Errors}

	When considering a simple on-axis two-mirror telescope, errors in positioning the secondary mirror result in aberrations
	described by a few low order Zernike polynomials.

	Off-axis designs may have different sensitivities, but their phase aberrations due to secondary mirror tilt, despace, and decenter
	are still described by combinations of low order polynomials.

\subsection{Sensitivity to stop size}

	The amount of energy leaking into the interior of the Lyot pupil is proportional to the
	fourth power of the bandwidth of the stop, or the inverse fourth power of the stop size
	expressed in resolution elements.  This follows from inspection of equation (\ref{Lpi_n}).
	The leaked energy finds is way into the final image plane, unless the Lyot stop is adjusted to block
	some of this light. This strong dependence must play a significant role when choosing a 
	coronagraph design given a desired contrast ratio derived from a scientific goal.

\subsection{Crosstalk between aberrations}

	When two aberration, either simple monomials or Zernike polynomials, are present simultaneously, the Lyot pupil
	field contains energy due to a `crosstalk' term in addition to effects from each individual term, even in the
	first order calculation.  Cataloguing the various crosstalk energy leaks is beyond the scope of this work: we restrict ourselves to
	a few simple, traditional aberrations.

\subsection{The effects of tilt and defocus}

	Equation (\ref{Lpi_n}) demonstrates that for a band-limited coronagraph
	there is complete cancellation of the central source even for small tilt 
	errors ($L_{\Pi,1} = 0$).  This particular
	result has already been derived by \citet{Kuchner02, Lloyd05}: here it is 
	derived as part of the band-limited coronagraph's response to a sequence of small polynomial phase errors.

	The way focus affects the field strength in the interior of the
	Lyot plane can be found by inspection of the $L_{\Pi,2}$ term in equation (\ref{Lpi_n}).
	Given an aberration of the the form $\alpha (x^2 + y^2)$, 
	the field strength in the interior of the Lyot plane is $\alpha d^2/6$, to first order.
	The interior of our coronagraph's Lyot plane fills uniformly with in-phase light.
	The effective telescope diameter describing the residual PSF for a
	circular aperture is $D - 2d$, so a fraction $\alpha^2 d^4 (1 - 2d/D)^2 / 36$
	of incident power leaks through the coronagraph.  It is distributed in
	an Airy pattern corresponding to an aperture $D - 2d$ in diameter, 
	assuming an optimally undersized Lyot stop is used.

	In the absence of other aberrations, this places a `ghostly PSF'
	with a calculable intensity in the final image. The form of this PSF for a circular unobstructed
	entrance aperture is just the Airy pattern of the undersized Lyot stop used in the coronagraph.
	This residual PSF is not to be confused with the spot of Arago \citep{Born93}, whose intensity varies with occulter size,
	depending on the relative size of an occulter in the pupil plane and the Fresnel zones it obscures.
	The spot of Arago's brightness fluctuates in intensity as one increases the pupil plane occulter size.
	The leaked PSF power we describe here does not possess undulating fluctuations, characteristic of the spot of Arago.
	The spot of Arago occurs with unaberrated wavefronts, whereas
	this leaked PSF is entirely dependent on the presence of aberrations in the wavefront.

	The analytical result presented here leaves no unresolved questions pertaining to the numerical accuracy or rounding error.
	The approximations used in deriving this result become more accurate as the magnitude of the phase aberrations decrease,
	so this result is pertinent to coronagraphs designed for very high Strehl ratio, high dynamic range applications. 
	The bandwidth of the focal plane mask function is the only coronagraphic parameter that enters into the light leak 
	through the coronagraph due to inexact focus in these regimes.

	We note in passing that if there is sufficient tilt error to merit a
	second order expansion of the pupil field strength,
	this second order tilt term will have a mathematically
	identical behavior to a small pure focus: a similar `ghostly' PSF will appear
	in the final coronagraphic image.  The fraction of incident power that
	leaks through the coronagraph can be calculated in a similar fashion to the way
	we estimate the power leak due to the first order effects of defocus.
	However, given the real or imaginary nature of the coefficients 
	in the expansion of the electric field in terms of the phase,
	the second order effect of tilt is in quadrature with the first order
	focus term (the former being purely imaginary, the latter real), so the two
	cannot be made to cancel each other: they will add power to
	the Lyot field (and the final coronagraphic PSF) in quadrature.

	If there is sufficient defocus to warrant a second order expansion
	of the pupil field, the amount of light leaking through
	the interior of the Lyot stop increases dramatically
	(Fig.~\ref{killerfig}, second row, second column).

	\subsection{Astigmatism}

	Small amounts of astigmatism have no first order effect on the band-limited coronagraph.  This is easy to understand 
	because of the separability of the $xy$ polynomial, and the nature of the convolution in the definition
	of equation (\ref{Lpi_n}).  The second order astigmatic power leak into the Lyot plane 
	is comparable to that of the second order tilt term, especially at the edges of the interior of the Lyot stop
	(Fig.~\ref{killerfig}, third column). 	Thus, off-axis telescope designs with residual astigmatism could
	be used with Lyot stops which are undersized relative to a Lyot stop tailored for perfect optics.  
	
	\subsection{Coma and spherical aberration}
	
	Leak through an unapodized Lyot coronagraph is dependent on local curvature of the wavefront, 
	so the first order leak due to these terms is high, especially at the edges of the Lyot pupil interior.
	First order light leak into the Lyot pupil interior due to spherical aberration has the same form as that of the
	second order leak due to defocus (Fig.~\ref{killerfig}), but, because of the relative quadrature of the first and second order
	field strengths, these aberrations cannot be used to cancel each other by design.

\section{Conclusion}

	Analytical studies of the mathematically simple band-limited coronagraph provide us with tools to help 
	evaluate coronagraphic behavior without the need for extensive numerical investigations.  The analysis also helps
	identify design strategies to reduce coronagraphic sensitivity to selected aberrations.  Our results suggest that
	off-axis unobscured high dynamic range apertures can be allowed more astigmatism
	if the Lyot pupil of a band-limited coronagraph is reduced in size, thereby  reducing the demands made upon the
	structural stability of the telescope.

	The theory developed here suggests that the coronagraphic PSF of the nominally on-axis source can be used to 
	determine best focus, by minimizing the on-axis intensity behind the focal plane mask, utilizing focus sweep data. 
	If the imaging quality is sufficiently high, leaked power due to small tilts (\ie\ small decentration of the target behind 
	the focal plane mask, rather than moving the target out from behind the spot) can be used to determine the location 
	of the focal plane mask. This latter exercise would aid astrometric calibration of the coronagraph.
	
	Further theoretical analyses of coronagraphic response to aberrations could assist coronagraph design and operational
	plans to be made, to improve the scientific productivity of high dynamic range space-based coronagraphs 
	dedicated to detecting and characterizing extrasolar planets.

\acknowledgments

The authors wish to thank the Space Telescope Science Institute's
Research Programs Office and its Director's Discretionary Research Fund.
We are grateful to P.~E.\ Hodge, J.~C.\ Hsu, P.~Greenfield, J.~T.\ Miller
and N.~Dencheva for their role in developing and supporting the
Python Numarray module \citep{numarray02, pycon03},
wrapping the numerical Fourier transform library FFTW \citep{fftw} for Numarray,
and providing support for matplotlib \citep{matplotlib}.
R.S.\ is supported by  NASA Michelson Postdoctoral Fellowship
under contract to the Jet Propulsion Laboratory (JPL) funded by NASA.
The JPL is managed for NASA by the California Institute of Technology, 
A.V.S.\ acknowledges support from Choate Rosemary Hall School,
and J.P.L.\ was supported in part by the California Institute of Technology's
Millikan fellowship. This work is based upon work supported by the National Science Foundation
under Grants No.\ AST-0215793,  AST-0334916 and has also been supported by 
the National Science Foundation Science and Technology Center for Adaptive Optics, managed by the
University of California at Santa Cruz under cooperative agreement No.\  AST-9876783.

\bibliographystyle{apj}
\bibliography{ms}

\begin{thebibliography}{31}
\expandafter\ifx\csname natexlab\endcsname\relax\def\natexlab#1{#1}\fi

\bibitem[{{Aime} \& {Soummer}(2002)}]{AS02b}
{Aime}, C.\ \& {Soummer}, R. 2002, in Astronomy With High Contrast Imaging: from
  planetary systems to avtive galactic nuclei, C.~Aime \& R.~Soummer Eds (EAS
  Publication Series)

\bibitem[{{Aime} \& {Soummer}(2003)}]{Aime03bb}
{Aime}, C.\ \& {Soummer}, R. 2003, EAS Publications Series, Volume~8,
  2003.~Astronomy with High Contrast Imaging, Proceedings of the conference
  held 13--16 May, 2002 in Nice, France.~Edited by C.~Aime and R.~Soummer, 8

\bibitem[{{Aime} {et~al.}(2002){Aime}, {Soummer}, \& {Ferrari}}]{ASF02}
{Aime}, C., {Soummer}, R., \& {Ferrari}, A. 2002, A\&A, 389, 334

\bibitem[{{Bloemhof} {et~al.}(2001){Bloemhof}, {Dekany}, {Troy}, \&
  {Oppenheimer}}]{Bloemhof01}
{Bloemhof}, E.~E., {Dekany}, R.~G., {Troy}, M., \& {Oppenheimer}, B.~R. 2001,
  \apjl, 558, L71

\bibitem[{Born \& Wolf(1993)}]{Born93}
Born, M.\ \& Wolf, E. 1993, Principles of Optics, 6th ed.\ (Cambridge: Cambridge
  University Press)

\bibitem[{{Breckinridge} \& {Oppenheimer}(2004)}]{Breckinridge04}
{Breckinridge}, J.~B.\ \& {Oppenheimer}, B.~R. 2004, \apj, 600, 1091

\bibitem[{{Digby} {et~al.}(2004){Digby}, {Oppenheimer}, {Newburgh}, {Brenner},
  {Shara}, {Mey}, {Mandeville}, {Makidon}, {Sivaramakrishnan}, {Soummer},
  {Graham}, {Kalas}, {Perrin}, {Roberts}, {Kuhn}, {Whitman}, \&
  {Lloyd}}]{DigbySPIE04}
{Digby}, A.~P., {Oppenheimer}, B.~R., {Newburgh},~L., {Brenner},~D., {Shara},~M.,
 {Mey},~J., {Mandeville},~C., {Makidon}, R.~B., {Sivara- makrishnan},~A.,
  {Soummer},~R., {Graham}, J.~R., {Kalas},~P., {Perrin}, M.~D., {Roberts},
  L.~C., {Kuhn},~J., {Whitman},~K., \& {Lloyd}, J.~P. 2004, in Proc.\ SPIE Vol.\ 
  5490, Advances in Adaptive Optics, Roberto Ragazzoni and Domenico Bonaccini;
  Eds.

\bibitem[{Frigo \& Johnson(1997)}]{fftw}
Frigo, M.\ \& Johnson, S.~G. 1997, in Technical Report MIT-LCS-TR-728
  (Massachusetts Institute of Technology)

\bibitem[{Green {et~al.}(2003)Green, Shaklan, \& Redding}]{Green03}
Green, J.~J., Shaklan, S.~A., \& Redding, D.~C. 2003, in Proc.\ SPIE, Vol.\ 4860,
  High-Contrast Imaging for Exo-planet Detection, ed.\ A.~B.\ Schultz \& R.~G.\ 
  Lyon

\bibitem[{Greenfield {et~al.}(2003)Greenfield, Miller, Hsu, \& White}]{pycon03}
Greenfield, P., Miller, J.~T., Hsu, J.-C., \& White, R.~L. 2003, in PyCon 2003
  Proceedings, ed.\ S.~Holden

\bibitem[{{Greenfield} {et~al.}(2002){Greenfield}, {Miller}, {Hsu}, \&
  {White}}]{numarray02}
{Greenfield}, P., {Miller}, T., {Hsu}, J.-C., \& {White}, R.~L. 2002, in ASP
  Conf.\ Ser.\ 281: Astronomical Data Analysis Software and Systems XI, 140--+

\bibitem[{Hunter(2005)}]{matplotlib}
Hunter, J. 2005, The Matplotlib User's Guide (University of Chicago Medical
  School)

\bibitem[{{Kasdin} {et~al.}(2003){Kasdin}, {Vanderbei}, {Spergel}, \&
  {Littman}}]{KVS03}
{Kasdin}, N.~J., {Vanderbei}, R.~J., {Spergel}, D.~N., \& {Littman}, M.~G.
  2003, ApJ, 582, 1147

\bibitem[{{Kuchner} \& {Traub}(2002)}]{Kuchner02}
{Kuchner}, M.~J.\ \& {Traub}, W.~A. 2002, \apj, 570, 900

\bibitem[{{Lloyd} \& {Sivaramakrishnan}(2005)}]{Lloyd05}
{Lloyd}, J.~P.\ \& {Sivaramakrishnan}, A. 2005, \apj, 621, 1153

\bibitem[{{Lyot}(1930)}]{Lyo30}
{Lyot}, B. 1930, C.~R.\ Acad.\ Sci Paris, 191, 834

\bibitem[{{Lyot}(1939)}]{Lyo39}
---. 1939, MNRAS, 99, 580

\bibitem[{{Mahajan}(1981)}]{Mahajan81}
{Mahajan}, V.~N. 1981, JOSA, 71, 75

\bibitem[{{Makidon} {et~al.}(2005){Makidon}, {Sivaramakrishnan}, {Perrin},
  i~{Roberts}, {Oppenheimer}, {Soummer,~R.}, \& {Graham}}]{MSP05}
{Makidon}, R.~B., {Sivaramakrishnan},~A., {Perrin}, M.~D., i~{Roberts}, L.~C.,
  {Oppenheimer}, B.~R., {Soummer,~R.}, \& {Graham}, J.~R. 2005, \pasp, {in
  press}

\bibitem[{{Malbet}(1996)}]{Malbet96}
{Malbet}, F. 1996, \aaps, 115, 161

\bibitem[{{Nisenson} \& {Papaliolios}(2001)}]{NP01}
{Nisenson}, P.\ \& {Papaliolios}, C. 2001, ApJL, 549

\bibitem[{{Noll}(1976)}]{Noll76}
{Noll}, R.~J. 1976, Optical Society of America Journal, 66, 207

\bibitem[{{Oppenheimer} {et~al.}(2004){Oppenheimer}, {Digby}, {Newburgh},
  {Brenner}, {Shara}, {Mey}, {Mandeville}, {Makidon}, {Sivaramakrishnan},
  {Soummer}, {Graham}, {Kalas}, {Perrin}, {Roberts}, {Kuhn}, {Whitman}, \&
  {Lloyd}}]{OppenheimerSPIE04}
{Oppenheimer}, B.~R., {Digby}, A.~P., {Newburgh},~L., {Brenner},~D., {Shara},~M., 
 {Mey},~J., {Mandeville},~C., {Makidon}, R.~B., {Sivaram- akrishnan},~A.,
  {Soummer},~R., {Graham}, J.~R., {Kalas},~P., {Perrin}, M.~D., {Roberts},
  L.~C., {Kuhn},~J., {Whitman},~K., \& {Lloyd}, J.~P. 2004, in Proc.\ SPIE Vol.\ 
  5490, Advances in Adaptive Optics, Roberto Ragazzoni and Domenico Bonaccini;
  Eds.

\bibitem[{{Oppenheimer} {et~al.}(2003){Oppenheimer}, {Sivaramakrishnan}, \&
  {Makidon}}]{OSM03}
{Oppenheimer}, B.~R., {Sivaramakrishnan},~A., \& {Makidon}, R.~B. 2003,
  {Imaging Exoplanets: The Role of Small Telescopes} (The Future of Small
  Telescopes In The New Millennium.\ Volume III~- Science in the Shadow of
  Giants), 155

\bibitem[{{Perrin} {et~al.}(2003){Perrin}, {Sivaramakrishnan}, {Makidon},
  {Oppenheimer}, \& {Graham}}]{Perrin03}
{Perrin}, M.~D., {Sivaramakrishnan}, A., {Makidon}, R.~B., {Oppen- heimer},
  B.~R., \& {Graham}, J.~R. 2003, \apj, 596, 702

\bibitem[{{Sivaramakrishnan} {et~al.}(2001){Sivaramakrishnan}, {Koresko},
  {Makidon}, {Berkefeld}, \& {Kuchner}}]{Sivaramakrishnan01}
{Sivaramakrishnan}, A., {Koresko}, C.~D., {Makidon}, R.~B., {Berkefeld}, T., \&
  {Kuchner}, M.~J. 2001, \apj, 552, 397

\bibitem[{{Sivaramakrishnan} \& {Lloyd}(2005)}]{SL05}
{Sivaramakrishnan}, A.\ \& {Lloyd}, J.~P. 2005, \apj, in press

\bibitem[{{Sivaramakrishnan} {et~al.}(2002){Sivaramakrishnan}, {Lloyd},
  {Hodge}, \& {Macintosh}}]{Sivaramakrishnan02}
{Sivaramakrishnan}, A., {Lloyd}, J.~P., {Hodge}, P.~E., \& {Macintosh}, B.~A.
  2002, \apjl, 581, L59

\bibitem[{{Soummer}(2005)}]{Soummer05}
{Soummer}, R. 2005, \apjl, 618, L161

\bibitem[{{Soummer} {et~al.}(2003{\natexlab{a}}){Soummer}, {Aime}, \&
  {Falloon}}]{SAF03}
{Soummer}, R., {Aime}, C., \& {Falloon},~P. 2003{\natexlab{a}}, A\&A, 397, 1161

\bibitem[{{Soummer} {et~al.}(2003{\natexlab{b}}){Soummer}, {Dohlen}, \&
  {Aime}}]{SDA03}
{Soummer}, R., {Dohlen}, K., \& {Aime},~C. 2003{\natexlab{b}}, A\&A, 403, 369

\end{thebibliography}

\begin{deluxetable}{lccccccccccc}
\tablecolumns{12}
\tablewidth{0pt}
\tablenum{1}
\footnotesize
\tablecaption{First and second order leak vs.\ Strehl ratio\label{table1}}
\tablehead{ 
\colhead{} & \multicolumn{2}{c}{99\% Strehl} & 
\colhead{} & \multicolumn{2}{c}{98\% Strehl} & 
\colhead{} & \multicolumn{2}{c}{95\% Strehl} \\
\cline{2-3} \cline{5-6} \cline{8-9}  \\ 
\colhead{} & \colhead{1$^{st}$ Order} & \colhead{2$^{nd}$ Order} & 
\colhead{} & \colhead{1$^{st}$ Order} & \colhead{2$^{nd}$ Order} & 
\colhead{} & \colhead{1$^{st}$ Order} & \colhead{2$^{nd}$ Order} \\
\colhead{} & \colhead{Power } & \colhead{Power } & 
\colhead{} & \colhead{Power } & \colhead{Power } & 
\colhead{} & \colhead{Power } & \colhead{Power }}
\startdata 
   Tilt  &  $                0 $ & $ 5.8\times 10^{-8}$  &&  $                0 $ & $  2.3\times 10^{-7}$  &&  $                0 $ & $ 1.5\times 10^{-6}$ \\
Focus  &  $1.8\times 10^{-5} $ & $ 4.9\times 10^{-8}$  &&  $3.6\times 10^{-5} $ & $ 2.0\times 10^{-7}$  &&  $9.1\times 10^{-5} $ & $ 1.3\times 10^{-6}$ \\
  Astig  &  $                0 $ & $ 2.3\times 10^{-9}$  &&  $                0 $ & $ 9.4\times 10^{-9}$  &&  $                0 $ & $ 6.1\times 10^{-8}$ \\
   Coma  &  $6.7\times 10^{-6} $ & $ 5.0\times 10^{-9}$  &&  $1.4\times 10^{-5} $ & $ 2.0\times 10^{-8}$  &&  $3.4\times 10^{-5} $ & $ 1.3\times 10^{-7}$ \\
 Sphere  &  $5.8\times 10^{-5} $ & $ 3.0\times 10^{-7}$  &&  $1.2\times 10^{-4} $ & $ 1.2\times 10^{-6}$  &&  $3.0\times 10^{-4} $ & $ 7.7\times 10^{-6}$ 
\enddata 
\end{deluxetable}

\end{document}